# Cold molecular gas in the Perseus cluster core

## Association with X-ray cavity, H$_\alpha$ filaments and cooling flow

Salomé, P.[1], Combes, F.[2], Edge A.C.[3], Crawford C.[4], Erlund M.[4], Fabian A.C.[4], Hatch N.A.[4], Johnstone R.M.[4], Sanders J.S.[4], and Wilman R.J[3]

[1] Institut de Radioastronomie Millimétrique, 300 rue de la Piscine, F-38406 St. Martin d'Hères, France e-mail: `salome@iram.fr`
[2] Observatoire de Paris, LERMA, 61 Av. de l'Observatoire, F-75014, Paris, France e-mail: `francoise.combes@obspm.fr`
[3] Department of Physics, University of Durham, South Road, Durham DH1 3LEi, UK e-mail: `alastair.edge@durham.ac.uk`
[4] Institute of Astronomy, Madingley Road, Cambridge CB3 OHA, UK

**Abstract.** Cold molecular gas has been recently detected in several cooling flow clusters of galaxies where huge optical nebulosities often stand. These optical filaments are tightly linked to the cooling flow and to the related phenomena, like the rising bubbles of relativistic plasma, fed by the radio jets. We present here a map in the CO(2-1) rotational line of the cold molecular gas associated with some H$\alpha$ filaments surrounding the central galaxy of the Perseus cluster: NGC 1275. The map, extending to about 50 kpc (135 arcsec) from the center of the galaxy, has been made with the 18-receiver array HERA, at the focus of the IRAM 30m telescope. Although most of the cold gas is concentrated to the center of the galaxy, the CO emission is also clearly associated to the extended filaments conspicuous in ionised gas and could trace a possible reservoir fueling the star formation there. Some of the CO emission is also found where the X-ray gas could cool down more efficiently: at the rims of the central X-ray cavity (where the hot gas is thought to have been pushed out and compressed by the central AGN expanding radio lobes). The CO global kinematics does not show any rotation in NGC 1275. The cold gas is probably a mixture of gas falling down on the central galaxy and of uplifted gas dragged out by a rising bubble in the intracluster medium. As recenlty suggested in other cluster cores, the cold gas peculiar morphology and kinematics argue for the picture of an intermittent cooling flow scenario where the central AGN plays an important role.

## 1. Introduction

Our view on cooling flows at the core of galaxy clusters has changed considerably in the recent years, due to X-ray observations with Chandra and XMM-Newton (Fabian et al. 2003). It has been confirmed that the hot X-ray gas, pervading the cluster, cools and condenses toward the centre of many galaxy clusters, but this new generation of satellites have also pointed out the complex morphology of the intracluster medium and suggested that the necessity of some re-heating mechanisms (Peterson et al. 2003).

At the same time, CO emission lines have been detected in several cooling flows at millimetre wavelengths, with the IRAM 30m telescope, the James Clerk Maxwell Telescope (JCMT) and the Caltech Submillimeter Observatory (CSO) (Edge 2001; Salomé & Combes 2003). For the first time, the presence of a very cold molecular gas within these environments has been revealed, and is now accepted. The Owens Valley Radio Observatory (OVRO) and the Plateau de Bure (PdB) interferometers have even produced the first maps of the molecular emission and confirmed the peculiar morphology and dynamics of the cold component (Edge & Frayer 2003; Salomé & Combes 2004a,b). The gas masses derived from these CO observations are also consistent with the cold residual gas expected to cool out of the X-ray band (from Chandra and XMM-Newton mass deposition rates), making it possible that the long searched for cool gas has indeed been detected.

The giant cD galaxy NGC 1275 lies in the center of the X-ray brightest cluster of galaxies in the sky: the Perseus cluster (Abell 426). This galaxy is at a redshift of 0.01756. At this distance, 1″ is 370 pc, with $H_0 = 70$ km.s$^{-1}$Mpc$^{-1}$. The gas in the core of this cluster will cool if there is no balancing heat source, due to the short cooling time of the intracluster medium (ICM) inside a few tens of kpc (Fabian et al. 2003).

The Perseus cluster (first detected in mm by Mirabel et al. 1989) remained the only cooling flow cluster core mapped in CO for about 10 years. During this time, different maps, lim-

*Send offprint requests to*: salome@iram.fr

ited in size and sensitivity, were performed, to probe the origin of the molecular emission. Reuter et al. (1993) imaged the millimetric emission of the central cluster region in CO(1-0) and CO(2-1) emission lines with the IRAM 30m telescope and built the first map of the central ∼50″. Braine et al. (1995) then observed the Perseus cluster in CO(1-0) with the IRAM Plateau de Bure interferometer (PdBI). Emission was detected around the nucleus but the continuum source made the map very noisy at the center. The authors suggested that the molecular gas could come from another source than a cooling flow like a recent merger event. More recently, Inoue et al. (1996a) observed the cluster center with the Nobeyama Millimeter Array in CO(1-0) in a primary beam of 65″ in diameter. Two peaks were identified within the inner 3 kpc (8 arcsec) which may be part of a ring like orbiting gas around the nucleus that could trace the AGN fueling of the cD galaxy. Finally, the most recent millimetric map of the central 1′ region has been made by Bridges & Irwin (1998) with the JCMT in CO(2-1) and CO(3-2) emission lines in single dish mode. Detections out to 36″ have been claimed, with a spatial resolution of 21″.

Large optical nebulosities are often observed within cooling flow clusters of galaxies (Crawford et al. 1999) and not detected in the surroundings of galaxies where the radiative cooling time is larger than the age of the cluster. The Perseus cluster harbors such a huge $H_\alpha$ filamentary nebulosity (Hu et al. 1983; McNamara et al. 1996a; Conselice et al. 2001). The origin of the optical filaments and their ionization source is not identified yet. However they do trace, in some part, the radiative cooling of the hot intracluster medium (directly/undirectly). Fabian et al. (2003) compared the X-ray structures with the optical emissions and proposed that the $H_\alpha$ filaments could be ionized cold gas that has been drawn up behind a rising bubble of cold gas, in a picture that takes into account the role of a central radio source in cooling flows (see also Boehringer et al. 1993).

Very recently, it has been found that the strong $H_\alpha$ emission in cooling flows also traces the presence of cold molecular gas. A strong correlation of $H_\alpha$ with CO, at ∼10-100 K (Edge 2001; Salomé & Combes 2003) has been discovered, which is reinforced by a clear association, both in morphology and dynamics, revealed in the IRAM PdBI CO(1-0) and CO(2-1) maps of Abell 1795. The $H_\alpha$ also correlates with the presence of warm $H_2$, at ∼1000-2000 K (Edge et al. 2002; Wilman et al. 2002). Recent United Kingdom Infra-Red Telescope (UKIRT) observations (Hatch et al. 2005a) have also shown the direct association of this $H_2$ with the outer optical filaments in NGC 1275. Based on near-IR Integral Field Unit (IFU) observations of the warm $H_2$, Wilman et al. (2005) have found a 50 pc radius ring in the very central part of the galaxy.

To probe the link between the molecular gas and the optical filaments within cooling flows, we undertook a large map of the central region of NGC 1275 in CO(2-1) emission. The next section present the observations procedure and the data reduction. In sections 3 and 4, the molecular gas emission detected with HERA on the 30m telescope is presented. The implication of this new view of the Perseus cluster core through millimetric wavelength is then discussed in section 5, before the conclusions in section 6.

**Table 1.** Parameters for NGC 1275 (at the distance of 77 Mpc adopted for NGC 1275, 1 arcsec= 0.37 kpc)

| Source | RA (J2000.0) | DEC (J2000.0) | $V_{Hel}$ (km/s) | Frequency* GHz |
|---|---|---|---|---|
| NGC 1275 | 03:19:48.16 | +41:30:42.1 | +5264 | 226.560 |

*Tuning frequency at CO(2-1)

## 2. Observations and Data reduction

The observations were made from 1st to 3rd January 2005 at the IRAM-30m telescope. We used the HEterodyne Receiver Array HERA, see Schuster et al. (2004), a focal array of 18 SIS receivers, 9 for each polarization, tuned at the CO(2-1) line, for NGC 1275 (226.56 GHz). The 9 pixels are arranged in the form of a center-filled square and are separated by 24″. The sampling was 6 arcsec (full sampling), and an homogeneous mapping procedure was used to regularly sweep a 12x12 pixels map, filling the intrinsic square, of 66x66″. Four such maps were done, covering the central 138x138″, and also a 5th one covering 66x66″ toward the north, centered at (0,108), over the northern vertical $H_\alpha$ filament. In total the map includes 720 points. The parameters for NGC 1275 are summarized in Table 1.

At 226 GHz, the telescope half-power beam width is 12″. The main-beam efficiency is $\eta_{mb} = T_A^*/T_{mb}$=0.57. The typical system temperature varied between 250 and 650 K (on the $T_A^*$ scale). Wobbler switching mode was used, with reference positions offset by 4′ in azimuth. The pointing was regularly checked on NGC1275 itself (3C84 continuum source) and the accuracy was 3″ rms. The backend used was WILMA providing a band of 1 GHz wide for each of the 18 detectors. The bands have 512 spectral channels spaced out by 2 MHz. The total bandwidth corresponds to 1300km/s at the CO(2-1) line (with velocity resolution of 2.6km/s).

The data were reduced with the GILDAS software. Some mis-functioning pixels were completely rejected (3 out of 18). Some spectra with random highly non-linear baselines were suppressed now and then. Linear baselines were subtracted from all other spectra, but the continuum at the center was impossible to detect, because of varying level (likely due to to varying atmosphere). The final spectra were smoothed to 30km/s.

## 3. Morphology of the cold molecular gas

### 3.1. East-West emission around the central cD galaxy

The major part of the CO emission comes from the central region. Some of the 720 spectra (the central ones) are displayed in Figure 1. The center of Perseus is clearly detected, and corresponds to the maximum of CO(2-1) emission. There is an offset with respect to the AGN center: the bulk of the molecular gas being shifted toward the West, by ∼ 3 kpc (8 arcsec). In addition, there is a clear detection of CO gas associated with the $H_\alpha$ emission toward the East and the West of the optical galaxy, with a total extent of about 30 kpc (80 arcsec). No emission is detected in the Eastern region by Inoue et al. (1996b) in CO(1-0), probably because the emission is resolved out by their interferometer (the primary beam is one arcmin). The total integrated emission is plotted on Figure 2.

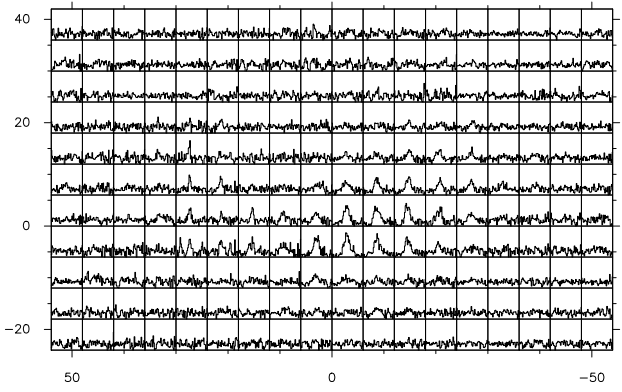

**Fig. 1.** Map of the central spectra taken toward NGC 1275. Each spectrum has a velocity scale from -600 to 600 km/s, and a temperature scale in $T_A^*$ from -10 to 40 mK. The position scale is RA-DEC in arcsec, centered on the 3C84 radio source.

### 3.2. Filamentary emission out to very large radii

The cold molecular gas is also detected all around the central 30 kpc (80 arcsec). The emission is fainter than in the centre, but still follow the filamentary $H_\alpha$ emission. We have extracted 115 spectra with S/N $\geq 4$ being particularly careful to broad emission lines (width $\leq 300$ km/s). Figure 3 shows the positions of CO(2-1) selected spectra compared with the regions where Conselice et al. (2001) extracted spectra of the ionized gas. A summary of the CO emission line parameters is given in Tab 3, 4 in the Appendix. The cold gas is still detectable between 20 and 50 kpc (50-120 arcseconds) away from the central galaxy, where, the cooling time of the X-ray ICM is still low: $2\text{-}3.10^8$ yr (Sanders et al. 2004).

### 3.3. Cold molecular mass distribution

We have computed the total molecular gas that can be deduced from the integrated CO emission, with simple assumptions about its excitation and metallicity. We have assumed that the antenna temperature in CO(2-1) is on average 0.7 times that in the CO(1-0) line, over the whole surface of the emission. In the center, the CO(2-1)/CO(1-0) ratio is equal to 1, and then decreases towards the outer parts down to 0.5, according to Reuter et al. (1993). Bridges & Irwin (1998) measured an average ratio of 0.74. Then we adopt the standard CO to $H_2$ conversion ratio, which should apply to solar metallicity gas, of $2.3 \times 10^{20}$ cm$^{-2}$ (Solomon et al. 1997).

We have evaluated the total mass by adding the contribution of all the different regions listed in Tab 3, 4 (Appendix) and taking into account overlapping beams. The total mass found is $4 \times 10^{10} M_\odot$, quite a large amount for a single galaxy. This is only a lower limit, as the gas coming from the ICM should have low metallicity. We have plotted, on Figure 4, the local and accumulated molecular gas mass versus radius. The present results are compatible with that from Reuter et al. (1993) and Bridges & Irwin (1998), who found a molecular gas mass close to the accumulated mass deduced here for a comparable radius. We find here quite a large mass of gas, very extended, showing that there is a lot of cold gas accompanying the filaments. The older measurements from Lazareff et al. (1989) and Mirabel et al. (1989) are slightly below the mass value we find with HERA.

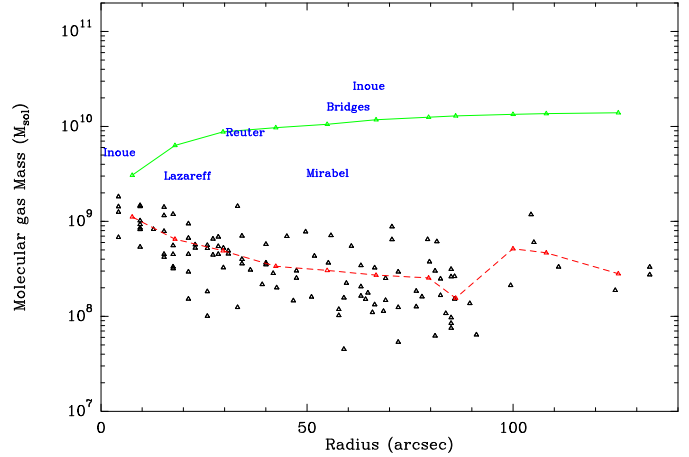

**Fig. 4.** Cold molecular gas mass evaluated from each CO(2-1) fitted line versus radius. The mean mass per point at each radius is overlaid in a dashed line. The continuous line represents the total accumulated mass at radius $\leq r$. Added are previous measurements with the author name.

The amount of cold gas in the core of the Perseus cluster is in agreement with the quantity of cooled residual gas expected from recent cooling rates of the intra-cluster medium. Bregman et al. (2005) deduced a mass deposition rate of 50 $M_\odot$/yr from OVI emission detected with FUSE in the $\sim 11$ kpc (30 arcsec) central region and X-ray data lead to $\sim 20$ $M_\odot$/yr (Fabian et al. 2005) in the same region. So the molecular gas detected here could have been accumulated in $\sim 1 \times 10^9$ yr, that is 3-4 cooling times in that region.

## 4. Dynamics of the cold molecular gas

### 4.1. Absence of rotation in the cD potential well

The kinematics deduced from the CO spectra is not regular, and it is not possible to follow a rotational pattern, as hinted at by previous observations (Reuter et al. 1993). Relative to the systemic velocity, there are negative velocities on both sides (West and East) on the major axis of the emission, and positive velocities in the center (see the isovelocity curves from Fig 5).

The average of all the selected spectra, plotted on Fig. 6, shows that the total emission cannot be fitted properly by a single gaussian. A two component model gives better results.

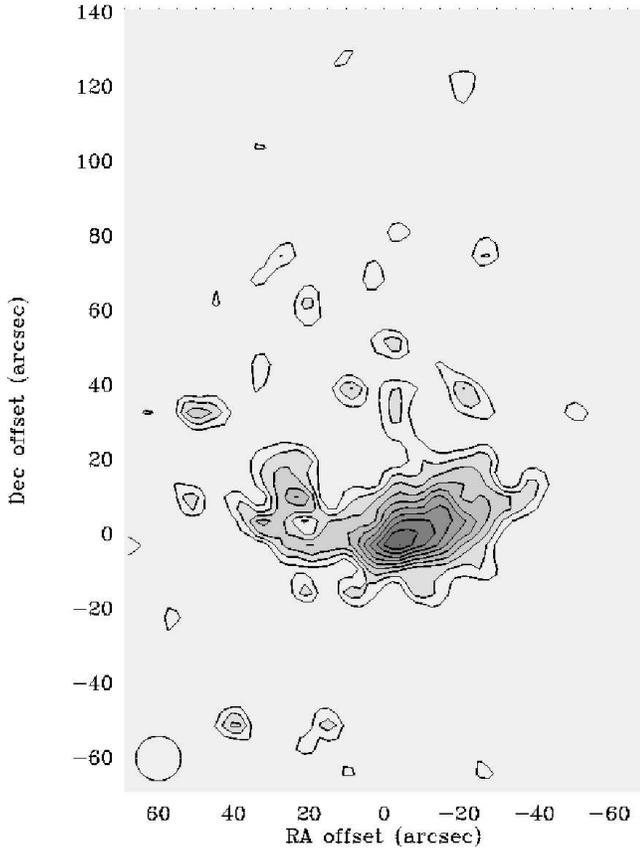 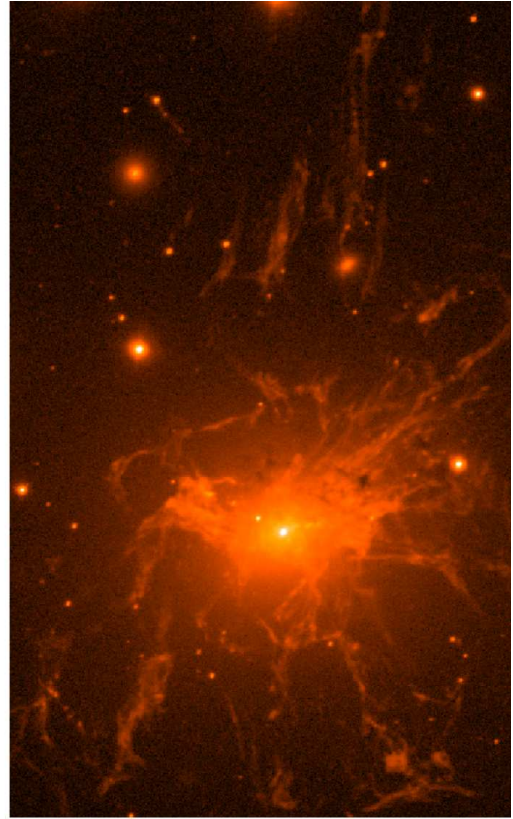

**Fig. 2.** On the left hand side: Integrated emission in CO(2-1) over the whole map. The region covered by the HERA observations is a central 138x138" map, with a northern 66x66" addition centered in (0,108") to cover the vertical north filaments. Contours are linear, form 10 to 100% of the maximum emission of 8.3 K km/s, in $T_A^*$ scale. The beam of 12" is indicated at the bottom left. On the right hand side: $H_\alpha$ image same scale, by Conselice et al. (2001)

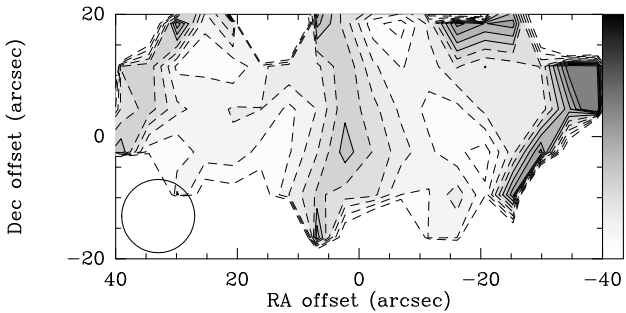 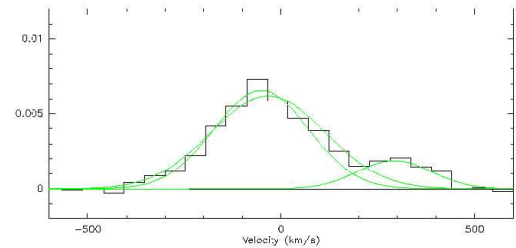

**Fig. 5.** Isovelocity map of the CO(2-1) emission. The contours are form -100 to 80km/s, by 20km/s. The beam of 12" is indicated at the bottom left. Negative velocities correspond to dashed contours.

**Fig. 6.** Average of the selected spectra plotted in Figure 3 ($T_A$ (K).vs. Velocity). Overlaid are 2 gaussian profiles: Positions are respectively 292.6 km/s and -51.3 km/s and widths are 226 km/s and 286 km/s.

We computed the CO velocity as a function of radius as shown on Fig. 7. Each position in this diagram has been identified with a number as referenced in Tab 3, 4. The velocity of the $H\alpha$ gas, computed by Conselice et al. (2001), has been added in red and cover the central 13 kpc (35 arcsec) region with a large scatter.

A comparison with a typical rotation curve expected for a spherical mass model for a galaxy like NGC 1275 shows that there is no clear sign of any rotating pattern of the CO gas.

Points are distributed all over the bound region. The cold gas between 18.5 and 37 kpc (50-100 arcsec) is detected at a velocity around 200 km/s in the cD rest frame. If the CO is coming from a cooling flow, we expect the gas to cool down in the cluster rest frame before being accreted by the cD galaxy. The cluster redshift is 0.0183 which represents +220 km/s in the cD rest frame. On Fig. 7, we have separated the points at a radius below 13 kpc (35 arcsec) from the points at a radius above this arbitrary limit (horizontal dashed line). In the filament, the CO clouds velocities stand between the cluster and the central

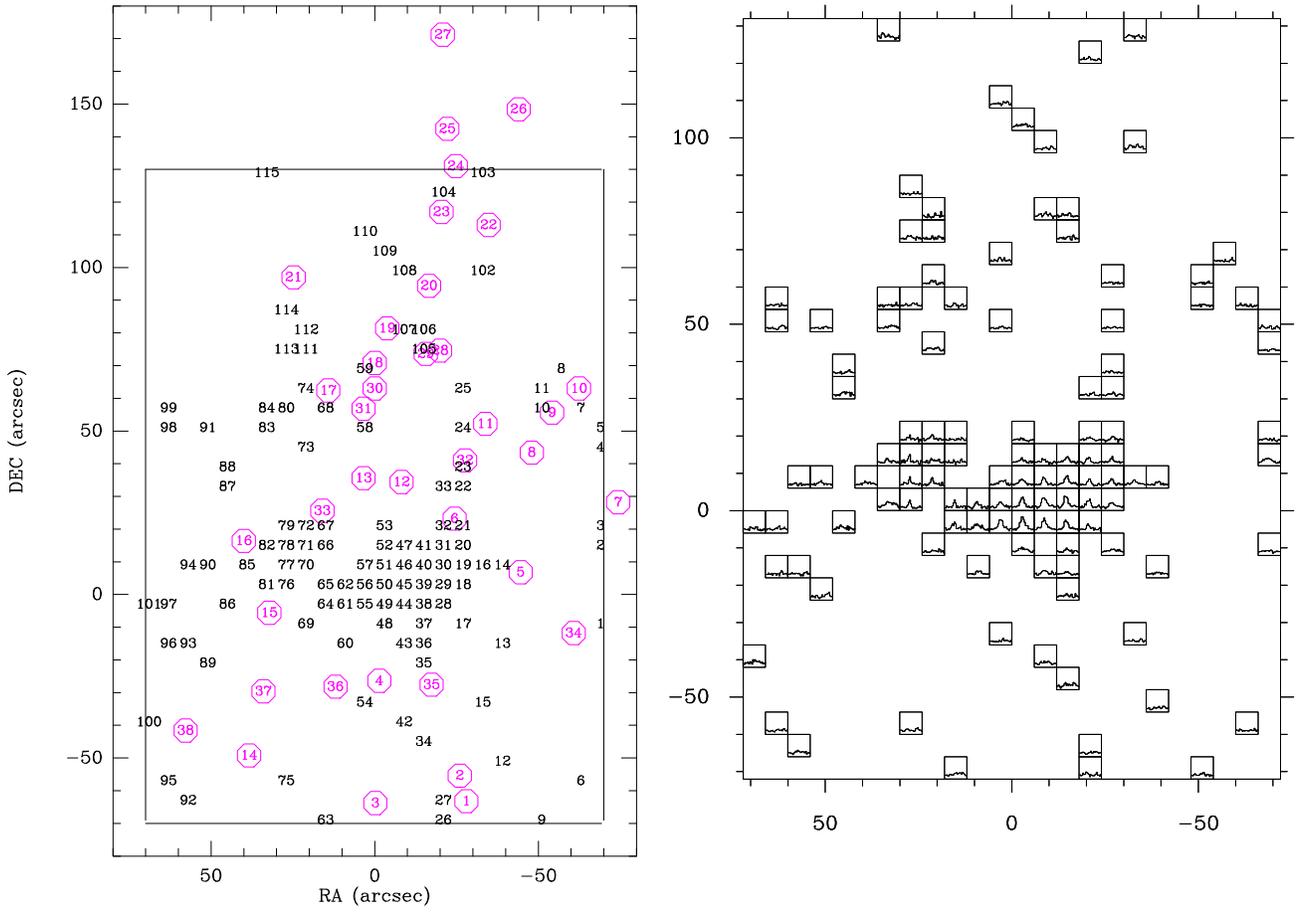

**Fig. 3.** On the left hand side: Positions of selected CO(2-1) spectra (in black) together with the $H_\alpha$ regions (in circles) observed by Conselice et al. (2001). Black numbers refer to regions described in Table 3. Circled red numbers refer to positions of Conselice's Fig. 5 regions. On the right hand side, the brightest CO(2-1) spectra all along the $H_\alpha$ filaments.

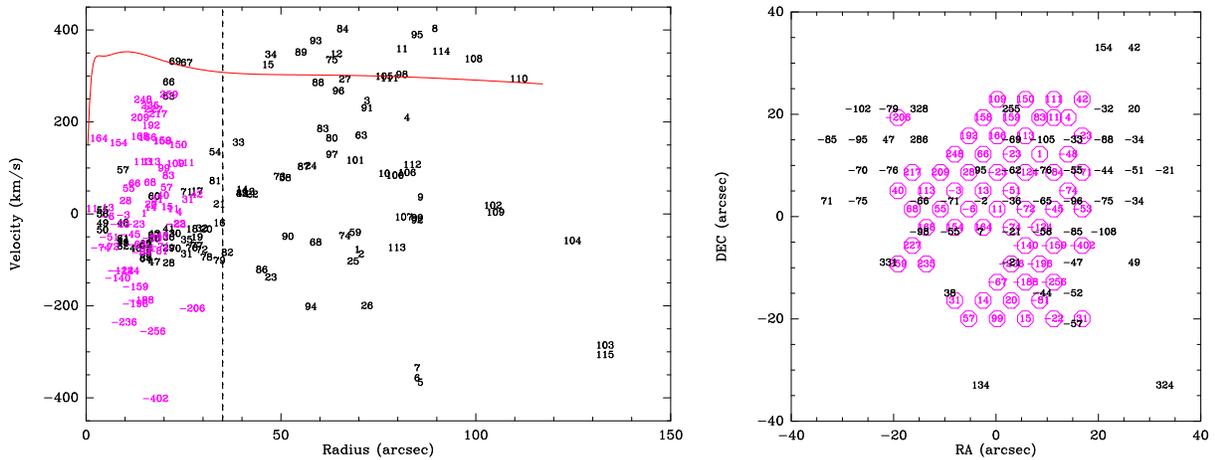

**Fig. 7.** On the left hand side - In black: CO(2-1) velocities versus radius. In purple: $H_\alpha$ velocity distribution from Conselice et al (2001). Overlaid in red: rotation curve expected for gas bounded in a galactic potential created by a bulge ($M_{bulge} = 3 \times 10^{10}$ $M_\odot$, $r_{bulge} = 0.6$ kpc), a disk ($M_{disk} = 1.4 \times 10^{11}$ $M_\odot$, $r_{disk} = 3.5$ kpc), a Black Hole ($M_{BH} = 4 \times 10^8$ $M_\odot$) and a Dark matter Halo ($M_{DM} = 5 \times 10^{11}$ $M_\odot$ in $r_{DM} = 42$ kpc, with then $M(r) \propto r^2$). On the right hand side - Comparison of the CO(2-1) velocity with the $H_\alpha$ filament velocity in the central region of Perseus.

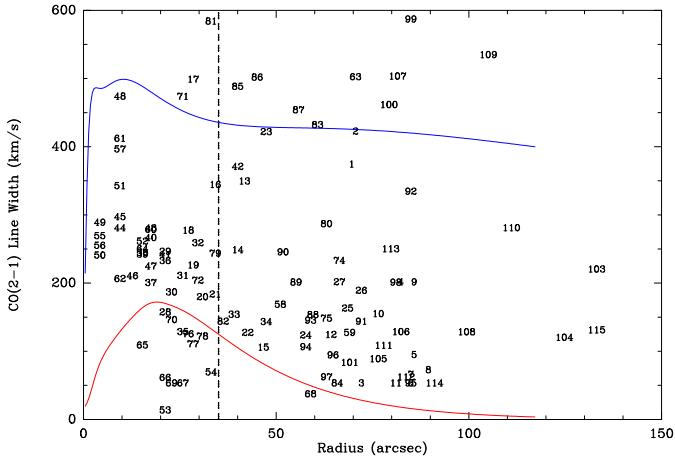

**Fig. 8.** CO(2-1) line width radial distribution. In blue: the escape velocity computed from the mass distribution detailed in Fig. 7. In red: $V_{crit} = 3.36G\Sigma/\kappa$: Toomre lower limit of the velocity dispersion for a rotating gas disk with $\Sigma = 500\,M_\odot.pc^{-2}$ to ensure the gas stability.

cD velocities. However, it is not possible to discern any clear velocity gradient here.

### 4.2. Two major components

Two different trends can be identified: close to the centre within 35", there is an offset ~ -150 km/s between the redshift of the optical galaxy and the CO rest frame. This offset might be an indication that the gas being accreted in the potential well of the central galaxy, is not yet completely relaxed. The Figure 7 compares the CO(2-1) and $H_\alpha$ kinematics. The velocities of the CO line are between 0 km/s and -100 km/s, while the $H_\alpha$ velocities are spread over +/- 250 km/s. Nevertheless most of the $H_\alpha$ regions detected close the CO regions also have negative velocities.

At larger radii, CO velocities show a larger scatter, but are in average in the positive sector. This component at large radii, associated with the long laminar $H_\alpha$ filaments, may trace the gas cooling down out of the hot ICM surrounding the cD galaxy. Such a cold gas resevoir in the filaments could fuel the star formation at large distances from the central galaxy that would help photoionize the surrounding gas.

This scenario is reinforced by the velocity dispersion measured on each of the lines. The Fig. 8 shows the CO(2-1) line widths versus radius. The data inside 13 kpc (35 arcsec) have a mean velocity dispersion around ~250 km/s, while in the outer region this value drops to ~125 km/s. This is still quite a large value. Some points even have very high velocity dispersion. These emitters could belong to cooling filaments expected to be dynamically perturbed and kinematically dissociated from the central cD. However, those high values could also be an effect of the larger data point spread where the emission is fainter (in particular in the outer regions).

The geometry of the source and the dynamical interaction between the ICM and the central radio lobes make the detailed interpretation more complex. It is possible, as suggested by Fabian et al. (2001), that part of the X-ray/$H_\alpha$ emission traces cooled uplifted gas, dragged behind the expanding radio lobes. Some of the X-ray excess is also found at the edges of the radio lobes. It is likely that some of the CO has formed in such cooler regions as the CO is expected to be on the form of dense clumpy clouds. The origin of gas detected here is certainly a mixture of different cooling processes inside a complex cooling flow scenario where the AGN plays a important role, re-heats the ICM, enhances the cooling along the radio edges, and drags some cooler gas (e.g. Crawford et al. (2005b), Crawford et al. (2005a)).

At ~ 30 kpc (80 arcsec) North, NGC 1275 harbors a long and thin optical filament extending radially North-South. Hatch et al. (2005b) determined the kinematics of the $H_\alpha$ and N[II] emission lines along this filament. The molecular gas, detected there, is found to share the same velocity structure (positions 105, 106, 102, 104, 103). The Northern and the Southern regions of the filament can be separated in two parts, flowing in opposite directions, as revealed through $H_\alpha$ spectroscopy by Hatch et al. (2005b). As suggested by the previous authors, this hint of gas not only falling down on the central galaxy, but also flowing away along the filament argues for the presence of a mechanism able to draw the gas away from the central cD. Whether the ICM gas is moving outward before it has cooled down to very low temperature (10-100 K) is still an open question. However, small dense clouds of molecular gas are very likely detected where they have formed if not perturbed by any external gravitational force. Whilst the central and eastern regions appear to be gas condensing and accumulating, do the extended regions represent an inflow or an outflow of cool gas ? In the scenario where the radio emitting plasma from the central AGN forms Buoyant bubbles rising into the ICM and dragging cool gas, could the cold material also be dragged out, therefore reducing the amount of accumulating mass on the central galaxy ? Comparison of $H_\alpha$ and CO kinematics along the so-called Horseshoe (filament extended Northwest of NGC 1275) cannot help to give a clear answer. While Hatch et al. (2005b) showed that this optical filament is most likely flowing out behind a rising bubble, the CO spectra are not sensitive enough to show that they also follow this dynamical model (positions 4, 10, 11 agree while positions 5, 7, 8 present some discrepencies). The molecular gas emission is very faint in those regions and it is hard to conclude from the present work. Deeper observations of those regions must now be achieved in millimeter to more accurately compare the optical and millimeter gas dynamics in the filaments.

## 5. Discussion

The present CO(2-1) observations allow us to understand the origin of the molecular gas in NGC 1275. The chaotic kinematics is not compatible with a disk in rotation, but supports the view of a system far from equilibrium, either due to a recent merger, or to gas accumulating from the cooling flow, and agitated by the radio-jets. The existence of young stellar clus-ters in the center, although NGC 1275 is an elliptical galaxy, is a consequence of this infall.

### 5.1. Tidal interaction or cooling flow

There are active star forming regions in the center of NGC 1275, revealed by the high resolution HST image (Holtzman et al. 1992; Carlson et al. 1998). The interpretation in terms of young globular clusters just formed in a recent merger is still debated. They could also come from the cooling flow, which is known to be intermittent, and can produce a starburst. There are stellar clusters associated with the high velocity (8200 km/s) system seen in absorption in front of the optical and X-ray emission, and also stellar clusters outside the optical galaxy, as far as 22 kpc (60 arcsec) from the center (Conselice et al. 2001). However, these outer stellar clusters do not coincide with the H$\alpha$ filaments, nor with the CO emission.

The merger hypothesis came from early optical spectroscopy, revealing two systems along the line of sight (Rubin et al. 1977): a smooth luminosity early-type system, i.e. NGC 1275 itself, at V=5200 km/s, seen unabsorbed in the south, and in front, superposed along the line of sight, a dusty late-type system, obscuring the first system in the North, at V=8200km/s, and showing a conspicuous NW extension. The detection of a broad absorption feature in HI (van Gorkom & Ekers 1983), but without any emission, did not support the existence of a late-type system. This high velocity component is seen by its optical emission lines, but no continuum, nor absorption lines are detected. The fact that no stellar component is detected was thought to be due to obscuration, due to the edge-on orientation of the possible late-type disk. However, near-infrared maps are available today to check this hypothesis, and no galaxy is found in the JHK map from 2MASS (Jarrett et al. 2003).

In their optical study, Rubin et al. (1977) and later Unger et al. (1990) found that the NW extension in the emission lines is seen in both velocities (5200 and 8200 km/s), and therefore support the view of an interaction between the two systems. Moreover, intermediate velocities have been observed in the recombining gas by Ferruit et al. (1997). It is interesting to note that the CO(1-0) map by Inoue et al. (1996b) found a North-West extension in the molecular gas also, at the low-velocity of 5200 km/s, and superposed on the optical extension. Since no galaxy is present in front of NGC1275, the high-velocity gas might be tidal debris or ram pressure debris likely to interact with the cooling flow, as proposed by Hu et al. (1983). Recently, Gillmon et al. (2004) showed that from X-ray absorption that the high velocity system is at least at 60 kpc (160 arcsec) out from NGC 1275 and not interacting yet.

The eastern part of the galaxy appears free from the high-velocity system, and the interpretation is easier. The H$\alpha$ filaments are all around NGC 1275, and appear tightly correlated with the X-ray bubbles (Fabian et al. 2003). The fact that the CO(2-1) emission is clearly associated to these H$\alpha$ filaments strongly supports the cooling flow hypothesis, for the origin of the cold molecular gas.

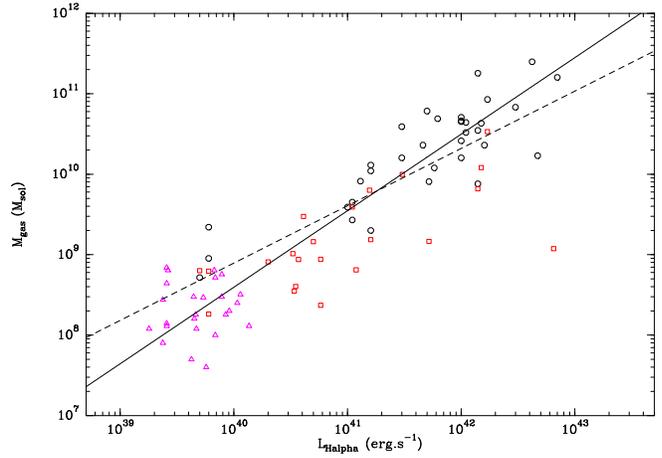

**Fig. 9.** Cold molecular gas mass as a function of H$\alpha$ luminosity for the regions in the filaments detected at both wavelength (see Tab 3). Circles are CO detections from Edge et al (2001), squares are CO detections from Salomé & Combes (2004a) and triangles are present results. Overlaid in dashed line is the linear relation fitted by Salomé & Combes (2003). Added in continuous line is a better relation which include low mass regions detected in NGC 1275 which also follow this relation.

### 5.2. Excitation mechanisms of the gas

The Figure 10 compares the CO intensity (Ico in K.km/s) and the H$_\alpha$ flux (erg.cm$^{-2}$.s$^{-1}$) radial distributions. We have normalized both emission lines by their respective maximum. The Ico decreases very steeply with radius. We have superposed $\propto$ r$^{-2}$ curves to mimic a central excitation source without any radiative loss. It is not clear whether we can rule out such a source of energy, since the points are highly scattered. Conselice et al. (2001) excluded the AGN as the main ionization source. So extra emission associated with star formation in the filaments are certainly involved, for example from shocks or UV radiation from young stars. We also compared the molecular gas mass to the H$\alpha$ luminosity for the individual regions detected in both lines, see Fig 9. We included the measurements obtained by Edge (2001) in this plot and Salomé & Combes (2003) for the ensemble of CO detected cooling flow clusters. The straight line overplotted is the linear relation fitted by Salomé & Combes (2003), pointed out by Edge (2001). Although NGC 1275 itself lies below the line in the Edge (2001) plot, the individual regions seem to follow it.

### 5.3. Intermittent cooling flow scenario

The CO contours appear to surround the northern cavity traced in the X-ray gas by the radio lobes (Figure 11). The hot gas, compressed towards the rims, cools there more efficiently, which could explain the presence of CO gas. This is similar to what Salomé & Combes (2004a) found, with the IRAM plateau de Bure interferometer, in another cooling flow cluster: Abell 1795. Like in Abell 1795, the H$_\alpha$ is enhanced along the radio lobes, where the CO is found. Surprisingly this correlation appears in both cases only along one of the lobes; moreover, active star formation is also identified in those regions. The present observations confirm what was found in Abell 1795: the radio lobe expansion, which is supposed to re-heat the ICM at large radius, can increase the cooling and accelerate the formation of cold molecular clouds along the radio lobes. Those cold clouds could then be accreted onto the central galaxy, or form stars directly along the radio lobe edges. More accurate estimate of the amount of gas available for the star formation compared to the amount of gas that can be accreted on the cD is crucial to constrain the intermittent cooling flow scenario. We notice that along the southern radio lobe, where no molecular gas is found, there is no tracer of star formation, nor is there clear $H_\alpha$ following the lobes. This reinforces the key role of molecular gas as a fuel for star formation traced by the $H_\alpha$ emission in filaments.

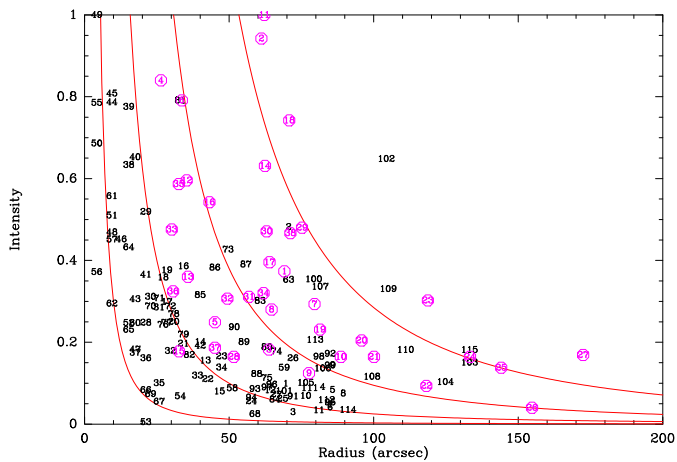

**Fig. 10.** Normalized CO(2-1) intensity (Ico in K.km/s) of selected spectra vs radius together with normalized $H_\alpha$ (erg.cm$^{-2}$.s$^{-1}$) flux from Conselice et al. 2001, cf Fig 3. Overlaid is a grid of $\propto r^{-2}$ curves.

### 5.4. Star formation regions inside a cooling flow

There is a clear filamentary and clumpy extension to the East (positions 70, 71, 72, 77, 78, 79, 82, 85), where Hatch et al. (2005a) have found $H_2$ ro-vibrational lines, and corresponding to intense filaments. The NW molecular gas region corresponds to the position of an identified star cluster found by Shields & Filippenko (1990).

The warm $H_2$ is not predominantly excited by stellar light in the outer filaments. Since excitation by the central AGN is insufficient, the most likely interpretation is a mixture between cooling gas, with thermal excitation, and with contributions of non-thermal excitation by stellar UV, X-rays, conduction from the ICM, or shocks (Hatch et al. 2005a).

The cold molecular gas is certainly fueling the star formation in that region. The Table 2 presents the parameters of the CO emission lines fitted in the Shields and Filippenko star cluster region (see Figure 12). Based on the Hatch et al. (2005a) results, we looked for two velocity components of the molecular gas emission in that region. We have fitted two Gaussian to the data and found that, in agreement with Hatch et al. (2005a), it is possible to distinguish a narrow line and a broad line (line 2 and line 1 respectively in Tab 2). The narrow line could trace the molecular gas reservoir associated to the star forming region and the broad line the underlying filament emission.

Shields & Filippenko (1990) found a total cluster mass of $5.10^6$ M$_\odot$ (assuming that the light is dominated by O-type stars) and reached an upper limit of $7.10^7$ M$_\odot$ with a steeper IMF. It represents a small amount compared to the $6.10^9$ M$_\odot$ of molecular gas found in that region, even with a very low star formation efficiency.

McNamara et al. (1996b) deduced from optical observations, that the star formation rate inside the central $\sim 15$ kpc (40 arcsec) can be around 40 M$_\odot$/yr during the past $7.10^7$ yr. The star formation rate is very close to the X-ray mass deposition rate in that region (20-50 M$_\odot$/yr). So the hot ICM cooling may steadily feed a reservoir of cold gas. If the star formation rate is slightly smaller than the mass deposition rate ($\sim 5$M$_\odot$/yr), then it is possible to accumulate a total mass of 2-5 $\times 10^{10}$ M$_\odot$ in 10 Gyr (as deduced from the present CO observations in the same region). It is also possible that the ICM only faces intermittent star formation events while the ICM cooling is continuous. Such an intermittent scenario could be due to the central AGN activity. If the star formation is triggered by shocks along the expanding radio lobes, this could occur in $10^7$ yr (McNamara et al. 1996b). Assuming that $\sim 100$ bursts of the AGN occurred in 10 Gyr. From Nulsen (2003), the energy necessary to re-heat the ICM and regulate the cooling flow needs an accretion of $\sim 10^6$M$_\odot$ by the central black hole. Over 10 Gyr the amount necessary to fuel the central black hole is still much smaller than the mass of molecular gas available, so only a small fraction has to be accreted by central galaxy.

## 6. Conclusions

The extended CO(2-1) map made with the HERA array at the IRAM-30m reveals a cold gas morphology very similar to the $H\alpha$ map from Conselice et al. (2001), in particular showing the same East-West extension in filaments. The eastern filaments are also conspicuous in warm $H_2$ emission (Hatch et al. 2005a). The CO contours are found surrounding the northern X-ray gas cavity formed by the radio jets from the central AGN. The CO kinematics does not show any rotation in the NGC 1275 galaxy.

The picture emerging from these observations is consistent with the interpretation that the hot intergalactic gas could have been pushed and compressed by the radio lobes in expansion. Along the lobes, where the gas is denser, the cooling is more efficient and the gas can cool down very quickly to very low temperatures. The molecular gas forms there, and maybe star formation can take place, providing some of the photoionisation to excite the $H\alpha$ emission.

*Acknowledgements.* Based on observations carried out with the IRAM 30m telescope. IRAM is supported by INSU/CNRS (France), MPG (Germany) and IGN (Spain). The authors also would like to acknowledge in particular the IRAM staff for help provided during the observations.

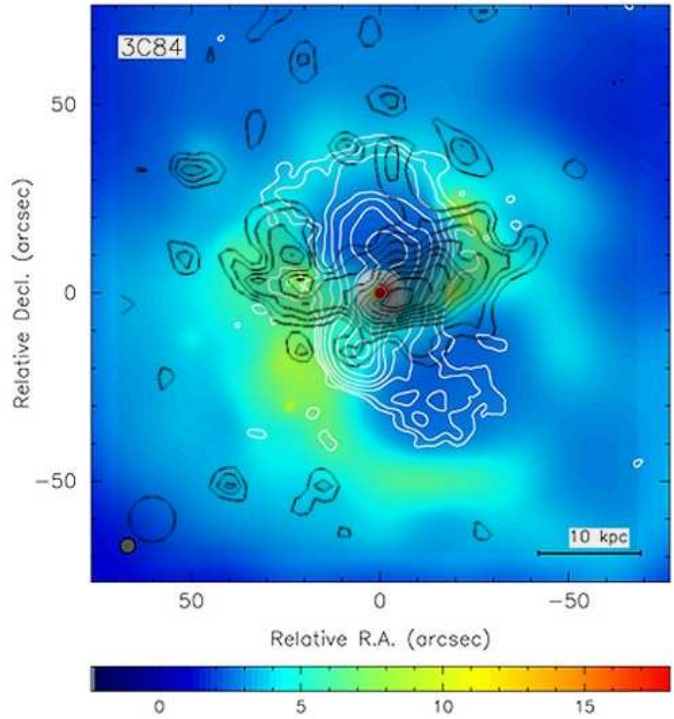

**Fig. 11.** Contours of the CO(2-1) emission (black) superimposed on the X-ray image (false colours, indicating intensity) by Fabian et al. (2003), and with the radio contours (white) by Pedlar et al. (1990). The relativistic plasma ejected by the central AGN in the two radio lobes pushes out the X-ray gas, which is compressed on the rims, and cools down there. The CO is also found at the border of the northern cavity.

| Position ("×") | Area 1 (K.km/s) | Area 2 (K.km/s) | Area2 / (Area1+Area2) |
|---|---|---|---|
| (21, 9)  | 2.14±0.8  | 1.76 ± 1.07 | 0.45 |
| (21, 15) | 0.88±0.36 | 2.14 ± 0.54 | 0.7  |
| (21, 21) | 1.33±1.01 | 1.05 ± 1.07 | 0.44 |
| (27, 9)  | 1.58±0.39 | 1.3 ± 0.73  | 0.45 |
| (27, 15) | 1.46±0.51 | 1.21 ± 0.73 | 0.45 |
| (27, 21) | 0.54±0.26 | 1.33 ± 0.59 | 0.7  |
| (33, 15) | 1.84±0.25 | 0.66 ± 0.22 | 0.26 |
| (39, 9)  | 1.13±0.18 | 1.02 ± 0.18 | 0.47 |
| All      | 1.4±0.22  | 1.3 ± 0.31  | 0.48 |

**Table 2.** Intensity of the CO(2-1) emission lines in the Shields and Filippenko star cluster region. Two Gaussians have been fitted. Around half of the total area is found in a broad line (300 km/s) and the other half is found in a narrower (130 km/s) component. The stronger emission could be associated to the star cluster and the smaller one to the underlying filament.

| Number | Position ("×") | Peak (mK) | S/N | Area (K.km/s) | Center (km/s) | Width (km/s) | $M_{gas}$ ($10^8 M_\odot$) | $F(H_\alpha)$ $10^{15}$erg.s$^{-1}$cm$^{-2}$ |
|---|---|---|---|---|---|---|---|---|
| 1 | (-70, -9) | 2.4 ± 0.4 | 5.9 | 0.9 ± 0.2 | -78.0 ± 58.0 | 374.0 ± 63.4 | 1.8 | 7.2 |
| 2 | (-70, 15) | 10.5 ± 1.3 | 7.5 | 4.7 ± 0.4 | -88.1 ± 20.1 | 422.1 ± 42.3 | 8.7 | |
| 3 | (-70, 21) | 5.1 ± 1.0 | 4.7 | 0.2 ± 0.2 | 245.6 ± 21.6 | 52.9 ± 59.7 | 0.5 | 6.6 |
| 4 | (-70, 45) | 4.1 ± 0.4 | 9.1 | 0.8 ± 0.2 | 209.2 ± 21.2 | 201.3 ± 39.1 | 1.6 | |
| 5 | (-70, 51) | 8.1 ± 1.8 | 4.4 | 0.8 ± 0.3 | -367.0 ± 18.1 | 94.5 ± 45.1 | 1.5 | |
| 6 | (-63, -58) | 7.1 ± 1.6 | 4.3 | 0.4 ± 0.1 | -357.3 ± 7.2 | 52.9 ± 132.0 | 0.7 | |
| 7 | (-63, 57) | 6.5 ± 0.9 | 6.9 | 0.4 ± 0.2 | -335.4 ± 22.0 | 65.6 ± 40.4 | 0.8 | 3.7 |
| 8 | (-58, 69) | 9.5 ± 0.9 | 10.0 | 0.7 ± 0.2 | 402.5 ± 8.4 | 72.3 ± 20.4 | 1.3 | |
| 9 | (-52, -70) | 6.6 ± 1.1 | 5.8 | 1.4 ± 0.3 | 35.7 ± 21.5 | 201.1 ± 48.7 | 2.6 | |
| 10 | (-52, 57) | 4.1 ± 0.8 | 5.0 | 0.6 ± 0.2 | 87.4 ± 28.6 | 154.6 ± 55.6 | 1.2 | 2.8 |
| 11 | (-52, 63) | 5.9 ± 0.9 | 6.5 | 0.3 ± 0.1 | 358.6 ± 24.0 | 52.9 ± 81.6 | 0.6 | |
| 12 | (-39, -52) | 6.1 ± 0.9 | 6.2 | 0.8 ± 0.3 | 347.4 ± 20.2 | 123.9 ± 38.9 | 1.5 | |
| 13 | (-39, -16) | 4.1 ± 1.0 | 4.0 | 1.5 ± 0.2 | 47.9 ± 23.8 | 349.1 ± 75.2 | 2.8 | |
| 14 | (-39, 9) | 7.4 ± 1.0 | 7.2 | 1.9 ± 0.3 | 52.7 ± 16.4 | 248.3 ± 35.2 | 3.6 | |
| 15 | (-33, -33) | 7 ± 1.6 | 4.3 | 0.7 ± 0.2 | 324.0 ± 11.5 | 105.4 ± 21.8 | 1.4 | |
| 16 | (-33, 9) | 10.3 ± 1.3 | 7.5 | 3.7 ± 0.5 | -20.0 ± 20.7 | 343.9 ± 74.4 | 7.0 | |
| 17 | (-28, -9) | 5.5 ± 1.0 | 5.4 | 2.9 ± 0.3 | 49.1 ± 30.0 | 498.4 ± 70.5 | 5.4 | |
| 18 | (-28, 3) | 11.9 ± 1.3 | 8.7 | 3.5 ± 0.3 | -33.3 ± 15.3 | 276.9 ± 32.4 | 6.5 | |
| 19 | (-28, 9) | 15.3 ± 1.4 | 10.7 | 3.6 ± 0.3 | -50.9 ± 11.0 | 225.9 ± 28.1 | 6.8 | |
| 20 | (-28, 15) | 12.8 ± 0.9 | 13.8 | 2.4 ± 0.2 | -33.8 ± 7.9 | 179.3 ± 20.4 | 4.5 | |
| 21 | (-28, 21) | 9.9 ± 0.6 | 16.1 | 1.9 ± 0.3 | 20.6 ± 16.8 | 183.1 ± 40.1 | 3.5 | 17.8 |
| 22 | (-28, 33) | 7.9 ± 0.9 | 8.8 | 1.0 ± 0.2 | 42 ± 16.0 | 126.8 ± 37.2 | 2 | |
| 23 | (-28, 39) | 3.6 ± 0.8 | 4.3 | 1.6 ± 0.2 | -138.4 ± 31.3 | 421.4 ± 70.5 | 3.0 | 6.9 |
| 24 | (-28, 51) | 4.1 ± 0.8 | 5.0 | 0.5 ± 0.2 | 104.1 ± 18.5 | 123.2 ± 35.8 | 1.0 | |
| 25 | (-28, 63) | 3.5 ± 0.8 | 4.0 | 0.6 ± 0.2 | -103.5 ± 30.2 | 162.6 ± 44.1 | 1.1 | |
| 26 | (-22, -70) | 7.8 ± 1.5 | 4.9 | 1.5 ± 0.5 | -200.2 ± 33.7 | 188.6 ± 93.5 | 2.94 | 8.4 |
| 27 | (-22, -63) | 3.3 ± 0.4 | 6.9 | 0.7 ± 0.3 | 292.4 ± 45.3 | 201.2 ± 136.5 | 1.3 | 21.2 |
| 28 | (-22, -4) | 14.5 ± 3.3 | 4.3 | 2.4 ± 0.4 | -107.3 ± 12.3 | 156.9 ± 29.9 | 4.5 | |
| 29 | (-22, 3) | 19.4± 2.9 | 6.6 | 5.0 ± 0.4 | -74.7 ± 10.0 | 245.9 ± 21 | 9.4 | |
| 30 | (-22, 9) | 15.4 ± 1.6 | 9.1 | 3.0 ± 0.3 | -43.1 ± 9.2 | 186.4 ± 20.0 | 5.6 | |
| 31 | (-22, 15) | 12.4 ± 1 | 12.3 | 2.7 ± 0.3 | -87.4 ± 13.6 | 210.6 ± 24.8 | 5.2 | |
| 32 | (-22, 21) | 6.3 ± 0.7 | 8.0 | 1.7 ± 0.3 | -31.8 ± 22.4 | 258.4 ± 45.7 | 3.2 | 17.8 |
| 33 | (-22, 33) | 7.1 ± 1.0 | 6.5 | 1.1 ± 0.3 | 154.9 ± 22.0 | 153.5 ± 52.8 | 2.1 | |
| 34 | (-16, -46) | 8.9 ± 1.4 | 6.0 | 1.3 ± 0.5 | 346.3 ± 24.4 | 142.8 ± 63.7 | 2.5 | |
| 35 | (-16, -22) | 7.2 ± 1.3 | 5.4 | 0.9 ± 0.4 | -56.9 ± 26.6 | 128.1 ± 76.1 | 1.8 | 13.2 |
| 36 | (-16, -16) | 6.3 ± 1.1 | 5.6 | 1.5 ± 0.2 | -51.0 ± 18.7 | 232.3 ± 35.8 | 2.9 | |
| 37 | (-16, -9) | 8.0 ± 1.6 | 4.7 | 1.7 ± 0.4 | -46.5 ± 22.5 | 200.1 ± 52.7 | 3.1 | |
| 38 | (-16, -4) | 23.9 ± 1.3 | 17.7 | 6.1 ± 0.4 | -84.3 ± 7.5 | 244.1 ± 17.8 | 11.56 | |
| 39 | (-16, 3) | 29.5 ± 2.1 | 13.7 | 7.5 ± 0.4 | -95.4 ± 6.2 | 241.8 ± 14.5 | 14.1 | |
| 40 | (-16, 9) | 22.6 ± 1.4 | 15.3 | 6.3 ± 0.5 | -54.0 ± 10.4 | 266.1 ± 29.2 | 11.9 | |
| 41 | (-16, 15) | 13.9 ± 1.7 | 7.8 | 3.5 ± 0.4 | -32.9 ± 12.0 | 241.2 ± 30.8 | 6.6 | |
| 42 | (-9, -39) | 4.7 ± 0.7 | 6.2 | 1.8 ± 0.3 | 44.0 ± 37.0 | 370.3 ± 61.2 | 3.5 | |
| 43 | (-10, -16) | 10.0 ± 2.3 | 4.3 | 2.9 ± 0.5 | -43.8 ± 25.4 | 280.3 ± 53.6 | 5.5 | |
| 44 | (-9, -4) | 25.8 ± 1.4 | 18.2 | 7.6 ± 0.4 | -57.8 ± 7.6 | 280.5 ± 21.7 | 14.3 | |
| 45 | (-10, 3) | 25.1 ± 1.9 | 13.1 | 7.9 ± 0.5 | -64.0 ± 9.8 | 296.8 ± 22.3 | 14.7 | |
| 46 | (-10, 9) | 19.8 ± 0.6 | 31.5 | 4.4 ± 0.2 | -75.1 ± 5.8 | 210.2 ± 13.1 | 8.2 | |
| 47 | (-10, 15) | 7.4 ± 1.2 | 6.0 | 1.7 ± 0.4 | -104.9 ± 27.1 | 224.5 ± 64.6 | 3.3 | |
| 48 | (-4, -9) | 9.1 ± 1.8 | 4.9 | 4.5 ± 0.4 | -20.0 ± 21.0 | 473.5 ± 43.5 | 8.5 | |
| 49 | (-4, -4) | 31.9 ± 1.4 | 21.5 | 9.7 ± 0.3 | -20.6 ± 4.4 | 288.5 ± 11.7 | 18.2 | |
| 50 | (-4, 3) | 26.3 ± 1.4 | 17.6 | 6.7 ± 0.2 | -35.3 ± 4.2 | 240.3 ± 9.9 | 12.5 | |
| 51 | (-4, 9) | 13.7 ± 1.4 | 9.5 | 4.9 ± 0.3 | -61.4 ± 10.0 | 342.2 ± 23.2 | 9.3 | |
| 52 | (-4, 15) | 8.7 ± 1.5 | 5.7 | 2.4 ± 0.3 | -68.0 ± 17.4 | 260.8 ± 48.5 | 4.5 | |
| 53 | (-4, 21) | 3.4 ± 0.8 | 4.0 | 0.1 ± 0.2 | 255.3 ± 20.92 | 13.2 ± 29.8 | 0.1 | |
| 54 | (3, -33) | 9.1 ± 1.4 | 6.2 | 0.6 ± 0.2 | 134.2 ± 10.8 | 69.0 ± 26.9 | 1.2 | 7.3 |
| 55 | (3, -4) | 26.9 ± 2.1 | 12.2 | 7.6 ± 0.3 | 7.3 ± 5.7 | 268.8 ± 15.2 | 14.3 | |
| 56 | (3, 3) | 13.4 ± 2.2 | 6 | 3.6 ± 0.4 | -1.1 ± 14.7 | 254.7 ± 33.8 | 6.7 | |
| 57 | (3, 9) | 10.5 ± 1.8 | 5.5 | 4.4 ± 0.4 | 95.0 ± 17.9 | 396.0± 39.0 | 8.2 | |
| 58 | (3, 51) | 4.8 ± 0.8 | 5.6 | 0.8 ± 0.2 | 78 ± 18.8 | 168.2 ± 48.1 | 1.6 | 7.0 |
| 59 | (3, 69) | 10.0 ± 2.1 | 4.6 | 1.3 ± 0.3 | -40.9 ± 14.8 | 126.9 ± 29.5 | 2.51 | 16.7 |
| 60 | (9, -16) | 8.1 ± 0.7 | 10.8 | 2.4 ± 0.1 | 38.0 ± 8.6 | 278.1 ± 16.4 | 4.5 | |

**Table 3.** Parameters of the brigtest CO(2-1) emission lines plotted in Fig. 3. The first and second columns give the region number as used in the related plots and the corresponding position offset in arcsec relative to the cD coordinates. The 3rd, 4th, 5th, 6th, 7th columns list the line parameters of fitted Gaussian profiles. The center of the lines is taken by comparison with the cD rest frame (z=0.01756). In column 8, we computed the cold molecular mass for each region, as explain in the text. Finally, the last column list the $H_\alpha$ flux in $10^{-15}$erg.s$^{-1}$.cm$^{-2}$ found by Conselice et al. (2001) in the region identified on the Figure 3.

| Number | Position ("×") | Peak (mK) | S/N | Area (K.km/s) | Center (km/s) | Width (km/s) | $M_{gas}$ ($10^8 M_\odot$) | $F(H_\alpha)$ $10^{-15}$erg.s$^{-1}$.cm$^{-2}$ |
|---|---|---|---|---|---|---|---|---|
| 61 | (9, -4) | 12.4 ± 1.7 | 7.0 | 5.4 ± 0.3 | -54.0 ± 16.3 | 411.6 ± 37 | 10.1 | |
| 62 | (9, 3) | 13.1 ± 2.9 | 4.4 | 2.8 ± 0.5 | -70.8 ± 21.1 | 205.9 ± 48.2 | 5.3 | |
| 63 | (15, -70) | 6.4 ± 1.2 | 5.0 | 3.4 ± 0.4 | 170.3 ± 31.5 | 501.9 ± 66.1 | 6.4 | 4.1 |
| 64 | (15, -4) | 15.9 ± 1.9 | 7.9 | 4.2 ± 0.5 | -97.8 ± 14.6 | 250.0 ± 34.2 | 7.8 | |
| 65 | (15, 3) | 19.6 ± 2.5 | 7.7 | 2.2 ± 0.2 | -65.6 ± 5.4 | 108.5 ± 11.4 | 4.2 | |
| 66 | (15, 15) | 12.6 ± 1.3 | 9.3 | 0.8 ± 0.2 | 286.4 ± 16.7 | 61 ± 179.0 | 1.5 | |
| 67 | (15, 21) | 9.6 ± 1.9 | 4.9 | 0.5 ± 0.2 | 328.4 ± 7.5 | 52.9 ± 761.6 | 1 | 10.7 |
| 68 | (15, 57) | 6.1 ± 1.4 | 4.6 | 0.2 ± 0.3 | -62.2 ± 19.0 | 37.0 ± 31.4 | 0.4 | 8.9 |
| 69 | (21, -9) | 12.8 ± 2.3 | 5.3 | 0.7 ± 0.2 | 331.2 ± 4.6 | 52.9 ± 179.6 | 1.3 | 4.0 |
| 70 | (21, 9) | 18.2 ± 3.2 | 5.6 | 2.8 ± 0.4 | -75.7 ± 9.8 | 145.8 ± 28.9 | 5.2 | |
| 71 | (21, 15) | 5.9 ± 1.4 | 4.1 | 3 ± 0.6 | 47.4 ± 40.8 | 473.2 ± 87.4 | 5.6 | |
| 72 | (21, 21) | 13.0 ± 2.2 | 5.7 | 2.8 ± 0.5 | -78.6 ± 14.1 | 203.5 ± 46.9 | 5.2 | 10.7 |
| 73 | (21, 45) | 5.5 ± 1.2 | 4.6 | 4.1 ± 0.8 | 80.7 ± 16.1 | 704.6 ± 135.3 | 7.7 | |
| 74 | (21, 63) | 7.0 ± 1.4 | 4.7 | 1.7 ± 0.3 | -48 ± 22.0 | 232.5 ± 37.5 | 3.2 | |
| 75 | (27, -58) | 7.0 ± 1.1 | 5.9 | 1.1 ± 0.2 | 334.5 ± 17.0 | 147.8 ± 32.1 | 2.0 | 14.2 |
| 76 | (27, 3) | 17.9 ± 1.5 | 11.2 | 2.3 ± 0.3 | -74.3 ± 7.2 | 124.9 ± 16.4 | 4.4 | 4.0 |
| 77 | (27, 9) | 20.7 ± 2.2 | 9.1 | 2.4 ± 0.4 | -69.1 ± 7.6 | 110.3 ± 21.7 | 4.5 | |
| 78 | (27, 15) | 20.4 ± 3.9 | 5.1 | 2.6 ± 0.4 | -94.6 ± 10.1 | 121.6 ± 24.8 | 4.9 | |
| 79 | (27, 21) | 8.3 ± 1.1 | 7.0 | 2.1 ± 0.5 | -101.8 ± 25.2 | 243.1 ± 75.0 | 4 | |
| 80 | (27, 57) | 6.0 ± 1.2 | 4.8 | 1.8 ± 0.4 | 164.3 ± 29.2 | 286.4 ± 61.9 | 3.4 | |
| 81 | (33, 3) | 12.5 ± 2.8 | 4.4 | 7.7 ± 0.7 | 71.4 ± 25.0 | 583.6 ± 54.1 | 14.4 | 4.0 |
| 82 | (33, 15) | 10.8 ± 1.6 | 6.7 | 1.6 ± 0.3 | -84.6 ± 12.9 | 143.4 ± 28.8 | 3.0 | 12.2 |
| 83 | (33, 51) | 6.4 ± 1.4 | 4.5 | 2.9 ± 0.2 | 184.8 ± 11.5 | 431.6 ± 26.8 | 5.5 | |
| 84 | (33, 57) | 10.4 ± 1.9 | 5.5 | 0.5 ± 0.2 | 401.7 ± 9.6 | 52.9 ± 187.6 | 1.0 | |
| 85 | (39, 9) | 5.9 ± 0.7 | 8.0 | 3.0 ± 0.5 | 44.1 ± 36.2 | 487.9 ± 81.6 | 5.7 | 12.2 |
| 86 | (45, -4) | 7.0 ± 1.4 | 4.2 | 3.7 ± 1.0 | -121.9 ± 61.7 | 501.4 ± 201.3 | 6.9 | 4.0 |
| 87 | (45, 33) | 7.9 ± 1.3 | 5.7 | 3.8 ± 0.3 | 102.2 ± 16.8 | 453.5 ± 33.0 | 7.1 | |
| 88 | (45, 39) | 7.3 ± 1.1 | 6.4 | 1.1 ± 0.3 | 285.0 ± 14.2 | 153.1 ± 47.2 | 2.2 | |
| 89 | (51, -22) | 9.2 ± 1.3 | 6.5 | 1.9 ± 0.4 | 351.0 ± 22.7 | 200.8 ± 46.6 | 3.6 | |
| 90 | (51, 9) | 8.9 ± 1.2 | 6.8 | 2.3 ± 0.3 | -49.5 ± 14.9 | 245.1 ± 38.6 | 4.3 | |
| 91 | (51, 51) | 4.3 ± 0.5 | 7.4 | 0.6 ± 0.2 | 229.8 ± 17.1 | 143.4 ± 31.8 | 1.2 | |
| 92 | (57, -63) | 4.7 ± 0.6 | 7.3 | 1.6 ± 0.5 | -13.9 ± 64.7 | 334 ± 128.5 | 3.1 | |
| 93 | (57, -16) | 5.6 ± 0.8 | 6.1 | 0.8 ± 0.3 | 376.0 ± 23.7 | 144.9 ± 44.7 | 1.5 | |
| 94 | (57, 9) | 5.6 ± 1.2 | 4.4 | 0.6 ± 0.2 | -202.0 ± 19.4 | 106.1 ± 39.5 | 1.1 | |
| 95 | (63, -58) | 9.2 ± 1.2 | 7.6 | 0.5 ± 0.1 | 389.1 ± 5.2 | 52.9 ± 583.8 | 0.9 | |
| 96 | (63, -16) | 9.4 ± 1.7 | 5.4 | 0.9 ± 0.2 | 267.0 ± 10.4 | 94.0 ± 18.5 | 1.7 | |
| 97 | (63, -4) | 13.3 ± 1.3 | 10.0 | 0.8 ± 0.2 | 128.7 ± 8.2 | 61.9 ± 55.4 | 1.6 | |
| 98 | (63, 51) | 7.5 ± 1.2 | 5.9 | 1.6 ± 0.3 | 302.6 ± 20.8 | 200.7 ± 38.8 | 3.0 | |
| 99 | (63, 57) | 2.2 ± 0.4 | 5.0 | 1.4 ± 0.2 | -8.9 ± 37.4 | 586.6 ± 83.1 | 2.6 | |
| 100 | (69, -39) | 7.0 ± 1.3 | 5.2 | 3.4 ± 0.5 | 83.3 ± 29.4 | 461.1 ± 73.4 | 6.4 | 10.5 |
| 101 | (69, -4) | 8.9 ± 1.3 | 6.3 | 0.7 ± 0.3 | 116.2 ± 15.3 | 83.1 ± 41.0 | 1.4 | |
| 102 | (-33, 99) | 8.6 ± 1.1 | 7.1 | 6.3 ± 0.9 | 17.8 ± 46.9 | 693.9 ± 99.6 | 11.8 | |
| 103 | (-33, 129) | 6.3 ± 1.4 | 4.2 | 1.4 ± 0.5 | -286.1 ± 38.2 | 220.0 ± 79.4 | 2.75 | 3.7 |
| 104 | (-21, 123) | 7.9 ± 1.3 | 5.8 | 1 ± 0.2 | -58.7 ± 14.0 | 119.8 ± 35.2 | 1.8 | |
| 105 | (-15, 75) | 10.5 ± 0.9 | 10.7 | 0.9 ±0.3 | 298.6 ± 15.4 | 88.4 ± 25.3 | 1.8 | |
| 106 | (-15, 81) | 9.8 ± 1.9 | 4.9 | 1.3 ± 0.3 | 89.2 ± 16.4 | 128.1 ± 37.5 | 2.4 | |
| 107 | (-10, 81) | 6.1 ± 0.3 | 15.9 | 3.2 ± 0.5 | -7.4 ± 45.1 | 502.9 ± 93.3 | 6.1 | |
| 108 | (-10, 99) | 8.3 ± 0.6 | 12.6 | 1.1 ± 0.23 | 336.7 ± 14.2 | 127.9 ± 29.1 | 2.1 | |
| 109 | (-3, 105) | 5.6 ± 1.0 | 5.6 | 3.2 ± 0.5 | 3.0 ± 43.1 | 534.4 ± 100.4 | 6.02 | |
| 110 | (3, 111) | 5.9 ± 1.4 | 4.2 | 1.7 ± 0.4 | 293.2 ± 28.2 | 280.6 ± 55.6 | 3.3 | |
| 111 | (20, 75) | 7.5± 1.6 | 4.6 | 0.8 ± 0.3 | 295.0 ± 26.6 | 108.2 ± 59.7 | 1.6 | |
| 112 | (20, 81) | 8.7 ± 1.2 | 7.0 | 0.5 ± 0.4 | 106.6 ± 33.1 | 61.7 ± 97.8 | 1.0 | |
| 113 | (26, 75) | 7.61 ± 1.3 | 5.8 | 2.0 ± 0.4 | -73.7 ± 23.5 | 249.6 ± 42.8 | 3.7 | |
| 114 | (26, 87) | 6.1± 1.4 | 4.3 | 0.3 ± 0.1 | 353.2 ± 19.0 | 52.9 ± 130.1 | 0.6 | |
| 115 | (33, 129) | 12.7 ± 2.8 | 4.4 | 1.7 ± 0.4 | -306.6 ± 14.9 | 130.7 ± 30.0 | 3.3 | |

**Table 4.** Tab 3 - continued

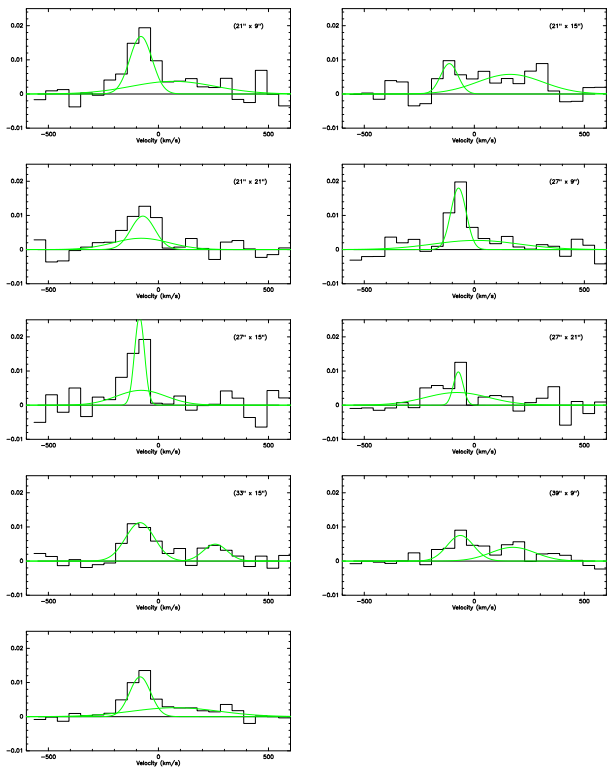

**Fig. 12.** Eight CO(2-1) spectra extracted from regions corresponding to the Shields and Filippenko star cluster region. The last spectrum at the bottom right is a line fitting over the mean emission over the 8 regions displayed here. When averaging all those spectra, two distinct components can be distinguished separated by ∼260 km/s: a broad line and a narrow line which could be coming from the filament and the star cluster region respectively.